\documentclass[12pt,letterpaper]{article}

\usepackage{amsmath,amssymb,calc}
\usepackage{graphicx}
\usepackage[nosort]{cite}

\makeatletter
\renewcommand\section{\@startsection {section}{1}{\z@}%
                                   {-3.5ex \@plus -1ex \@minus -.2ex}
                                   {2.3ex \@plus.2ex}%
                                   {\normalfont\large\bfseries}}
\renewcommand\subsection{\@startsection{subsection}{2}{\z@}%
                                     {-3.25ex\@plus -1ex \@minus -.2ex}%
                                     {1.5ex \@plus .2ex}%
                                     {\normalfont\bfseries}}
\makeatother

\let\non\nonumber

\let\a=\alpha

\let\s=\sigma

\newcommand{\del}{\partial}

\def\one{^{(1)}}

\newcommand{\bea}{\begin{eqnarray}}
\newcommand{\eea}{\end{eqnarray}}
\newcommand{\be}{\begin{equation}}
\newcommand{\ee}{\end{equation}}
\newcommand{\bma}{\begin{pmatrix}}
\newcommand{\ema}{\end{pmatrix}}

\newcommand{\hlf}{\frac{1}{2}}

\newcommand{\cM}{{\cal M}}

\newcommand{\La}{\Lambda}
\newcommand{\la}{\lambda}

\newcommand{\G}{\Gamma}
\newcommand{\e}{\epsilon}

\newcommand{\dd}{\delta}

\newcommand{\m}{\mu}
\newcommand{\n}{\nu}

\newcommand{\f}{\psi}

\newcommand{\tg}{\tilde g}
\newcommand{\hg}{\hat g}

\newcommand{\ap}{\alpha'}

\newcommand{\hn}{\hat{\nabla}}
\newcommand{\tn}{\tilde{\nabla}}
\newcommand{\dom}{\dot{\omega}}
\newcommand{\ddo}{\ddot{\omega}}
\newcommand{\da}{\dot{a}}
\newcommand{\dda}{\ddot{a}}

\newcommand{\C}[1]{$(\ref{#1})$}

\def\IZ{\relax\ifmmode\mathchoice
{\hbox{\cmss Z\kern-.4em Z}}{\hbox{\cmss Z\kern-.4em Z}}
{\lower.9pt\hbox{\cmsss Z\kern-.4em Z}} {\lower1.2pt\hbox{\cmsss
Z\kern-.4em Z}}\else{\cmss Z\kern-.4em Z}\fi}
\def\IR{\relax{\rm I\kern-.18em R}}

\def\one{{\hbox{ 1\kern-.8mm l}}}

\def\tr{{\rm tr\,}}

\newlength{\bredde}
\def\slash#1{\settowidth{\bredde}{$#1$}\ifmmode\,\raisebox{.15ex}{/}
\hspace*{-\bredde} #1\else$\,\raisebox{.15ex}{/}\hspace*{-\bredde}
#1$\fi}

\newsavebox{\zzzbar}
\sbox{\zzzbar}
  {\setlength{\unitlength}{0.9em}
  \begin{picture}(0.6,0.7)
  \thinlines
  \put(0,0){\line(1,0){0.6}}
  \put(0,0.75){\line(1,0){0.575}}
  \multiput(0,0)(0.0125,0.025){30}{\rule{0.3pt}{0.3pt}}
  \multiput(0.2,0)(0.0125,0.025){30}{\rule{0.3pt}{0.3pt}}
  \put(0,0.75){\line(0,-1){0.15}}
  \put(0.015,0.75){\line(0,-1){0.1}}
  \put(0.03,0.75){\line(0,-1){0.075}}
  \put(0.045,0.75){\line(0,-1){0.05}}
  \put(0.05,0.75){\line(0,-1){0.025}}
  \put(0.6,0){\line(0,1){0.15}}
  \put(0.585,0){\line(0,1){0.1}}
  \put(0.57,0){\line(0,1){0.075}}
  \put(0.555,0){\line(0,1){0.05}}
  \put(0.55,0){\line(0,1){0.025}}
  \end{picture}}

\newcommand{\ena}{\end{eqnarray}}
\newcommand{\beqa}{\begin{eqnarray}}
\newcommand{\eeqa}{\end{eqnarray}}

\newcommand{\half}{\frac{1}{2}}

\def\G{\Gamma}

\newfont{\goth}{ygoth.tfm scaled 1200}                   

\def\a{\alpha}

\def\e{\epsilon}

\def\f{\phi}

\def\m{\mu}
\def\n{\nu}

\def\s{\sigma}

\def\G{\Gamma}

 \numberwithin{equation}{section}

\def\1{{(1)}}
\def\2{{(2)}}
\def\3{{(3)}}

\def\1{{\bf 1}}
\def\a{{\alpha}}

\def\M{{\mathcal M}}
\def\al{{\alpha'}}

\begin{document}
\begin{titlepage}

\begin{center}

{February 13, 2012} \hfill         \phantom{xxx} \hfill EFI-11-25

\vskip 2 cm {\Large \bf Constraints on String Cosmology }
\vskip 1.25 cm {\bf  Stephen R. Green\footnote{srgreen@uchicago.edu}, Emil J. Martinec\footnote{ejmartin@uchicago.edu}, Callum Quigley\footnote{cquigley@uchicago.edu} and Savdeep Sethi\footnote{sethi@uchicago.edu}}\non\\
\vskip 0.2 cm
 {\it Enrico Fermi Institute, University of Chicago, Chicago, IL 60637, USA}\non\\ \vskip 0.2cm

\end{center}
\vskip 2 cm

\begin{abstract}
\baselineskip=18pt

String theory contains sources like orientifold planes that support higher derivative interactions. These interactions make possible static flux compactifications which are forbidden in supergravity. They can also lead to violations of the strong energy condition (SEC) which is needed for an accelerating universe. We examine how large a violation is possible in the context of the heterotic string compactified to four dimensions. We find that de Sitter solutions are still not possible but that classically forbidden anti-de Sitter solutions are possible.


\end{abstract}

\end{titlepage}


\section{Introduction}

Understanding accelerating universes remains one of the primary outstanding challenges for string theory. Acceleration requires a violation of the strong energy condition. However, the supergravity theories describing low-energy string theory in ten dimensions or M-theory in eleven dimensions satisfy the strong energy condition. This property is inherited on compactification ruling out accelerating solutions in supergravity~\cite{Gibbons:1984kp}. 

However, string theory is not supergravity. There are ingredients like orientifold planes in type II string theory which  support higher derivative interactions that can lead to violations of the strong energy condition (SEC). These ingredients are present in every corner of the string landscape: in M-theory, they come  from eight derivative modifications to eleven-dimensional supergravity. In the heterotic and type I strings, they come about from a class of four derivative interactions which are leading order in the $\a'$ expansion. 

It is precisely these interactions which lead to background charges on compact spaces, permitting stringy flux compactifications that evade the constraints of supergravity. It is important to note that the stringy interactions are crucial even for large volume compactifications, like those found in type IIB string theory, which are well described by supergravity. The background charges generated by the stringy interactions permit the solution of the supergravity equations of motion which involve Gauss Law type constraints.   

The first example of this kind appeared in the heterotic string~\cite{Strominger:1986uh}. The Bianchi identity for the heterotic string,
\be \label{bianchiintro}
d {H} = { \alpha' \over 4}  \left[\tr (R\wedge R) - \tr (F \wedge F)  \right],
\ee
contains a gravitational $\a'$ correction that induces background NS5-brane charge. The presence of this coupling makes possible supersymmetric compactifications with $H$-flux (torsional compactifications) to four-dimensional Minkowski space-time. There is a similar story for flux compactifications of M-theory to three dimensions~\cite{Becker:1996gj}, and type IIB compactifications to four dimensions~\cite{Dasgupta:1999ss}. 
  
The goal of this project is to analyze the effects of these intrinsically stringy ingredients with the aim of seeing whether these ingredients are sufficient to permit accelerating universes in string theory. The no-go theorem of~\cite{Gibbons:1984kp}, reviewed in~\cite{ Maldacena:2000mw, Gibbons:2003gb}, assumes no time-dependent scalar fields. In this work, we will also assume that there are no time-dependent scalar fields. Considering time-dependent scalar fields introduces many interesting model-dependent issues that will be discussed in a companion paper. 

In this work, we will arrive at a fairly model independent set of results. We will analyze heterotic string compactifications using the string effective action simply because this is the most tractable setting: the essential couplings for violating the supergravity no-go theorems already appear at tree-level in the ten-dimensional string effective action and are known. By contrast, considering type II compactifications would require introducing orientifold planes and adding their supported couplings to the bulk supergravity action. The complete set of supported orientifold or higher derivative couplings, unfortunately, is not yet known in type II string theory and M-theory. Fortunately, we expect to see the same qualitative physics found here for the heterotic string from other possible compactifications. That is certainly the case with Minkowski flux compactifications: the same qualitative phenomena are found in every flavor of compactification. 

For compactifications to Minkowski space-time, we find that the string frame metric is always unwarped, regardless of whether the compactification preserves supersymmetry. All the warping can be viewed as coming from the way the dilaton varies over the compactification space.  
We also find that de Sitter space solutions are robustly excluded, even when the higher derivative interactions are included. However, AdS$_4$ solutions are permitted because of the leading $\a'$ interactions. These solutions are not possible in heterotic supergravity. These AdS backgrounds are not supersymmetric, which is why they were not seen in the analysis of~\cite{Strominger:1986uh}.
It would be very interesting to find a world-sheet formulation for these compactifications. In terms of past literature, there have been interesting investigations of the kind of energy condition violation needed to obtain inflation from higher-dimensional theories in~\cite{Wesley:2008fg, Steinhardt:2008nk}. Inflation has also been explored from the world-sheet perspective in~\cite{Schalm:2010hb}. 

In terms of future work, we will explore similar constraints on dynamical scalars in a companion project. We should stress that the only quantum effects missed in this analysis are string non-perturbative effects. There are no string loop corrections to this order in the momentum expansion. Non-perturbative ingredients like heterotic five-branes are unlikely to directly alter our analysis since these branes can be viewed as the zero size limit of Yang-Mills instantons. The stress-energy contribution from instantons is fully included here. On the other hand, non-perturbative gauge dynamics (including gauge groups supported on branes) has at least the potential to significantly modify our conclusions. Another interesting extension would be the consideration of compactification manifolds with boundary, though little is known about the nature of boundaries in the heterotic string.   

It would also be interesting to explore what differences emerge from a similar analysis of type II and M-theory compactifications, assuming one can gain sufficient control over the higher derivative interactions. It would also be fascinating to derive analogous results directly from the string world-sheet without recourse to a derivative expansion of the string effective action. Past work on acceleration from gravity theories, some with higher order corrections, includes~\cite{Ohta:2003ie, Ohta:2004wk, Maeda:2004hu, Maeda:2004vm, Akune:2006dg, Bamba:2007ef}.


\section{The Heterotic Effective Action}
\subsection{Metric conventions}
\label{metricconventions}

The complete set of conventions used here is described in Appendix~\ref{conventions}, along with details of the curvature calculations.  
Consider compactifying $10$-dimensional space-time on a warped product space $X_d \times_W \M$. Choose a general warped metric,
\be\label{wmetric}
\widetilde{ds}^2 = W(y)^2 \left( \hg_{\mu\nu} dx^\mu dx^\nu + \hg_{mn} dy^m dy^n \right).
\ee
We use a tilde to denote the Einstein frame metric, and a hat to denote the unwarped, product frame metric. The string frame metric will remain undecorated. The relation between string, Einstein, and product frames is given by
\be\label{metricrel}
g_{MN} = e^{\f/2}\tilde{g}_{MN} = e^{\f/2}W^2\hg_{MN},
\ee
where $\f$ is the string dilaton. The warp factor and the dilaton naturally combine into a single conformal factor
\be
\omega = \log W + {1\over4}\f,
\ee
which we can use to go between string and product frames.
Typically in supersymmetric heterotic compactifications, the warp factor is determined only by the dilaton so the string-frame metric is unwarped. A priori, it is not clear that this is true for backgrounds that preserve no supersymmetry. 

Take $X_d$ to be a Lorentzian space-time of FLRW form with metric
\be
\widehat{ds}_X^2 = -dt^2  +a^2(t)h_{ij}dx^i dx^j.
\ee
We can specialize to cases with more isometries as needed. Assume the internal manifold $\M$ is compact without boundary. Generically, $\M$ will have no isometries.

\subsection{The ten-dimensional string effective action}

Let us start with the effective action. There is one length scale in the problem, which is the string length $\ell_s$ with $\al = \ell_s^2$. The field content consists of the metric $g$, dilaton $\phi$, NS $2$-form potential $B$  with three-form field strength $H$ defined below, and Yang-Mills gauge-fields $A$.  The bosonic couplings in the heterotic space-time effective action are given by
\begin{equation} \label{hetaction}
\begin{split}
S={1\over 2\kappa^2} \int d^{10}x \sqrt{-g} \, e^{-2 \phi}& \Big[R
+4 (\partial \phi)^2-{1\over 2} | {H} |^2 \cr & -
{\alpha' \over 4} \left( \tr | {F}|^2 -\tr |
R_+|^2\right)   +O(\alpha'^2) \Big],
\end{split}
\end{equation}
where
\begin{equation}
\tr | R_+|^2={1\over 2} R_{MNAB} (\Omega_+)R^{MNAB}
(\Omega_+)
\end{equation}
and $ {F}$ is the Yang-Mills field strength. This action is expressed in string frame, but it will be more convenient for checking energy conditions to conformally transform to Einstein frame in which the Einstein-Hilbert term is canonical.

The Einstein-Hilbert term is constructed using the standard metric
connection. The Riemann tensor appearing in the
$O(\alpha')$ correction is constructed using the connection $\Omega_+$
where
\begin{equation}\label{conn}
{\Omega^{AB}_\pm}_M= {\Omega^{AB}}_M\pm \half {{
H}^{AB}}_M+O(\alpha'),
\end{equation}
and ${\Omega}$ is the spin connection. The definition of ${ H}$ already
includes $O(\alpha')$ corrections,
\begin{equation}\label{aaav}
{H} = dB +\frac{\alpha'}{4}   \left[ {\rm CS}(\Omega_+) - {\rm
CS}(A)  \right] ,
\end{equation}
where $A$ is gauge field and CS denotes the Chern-Simons invariant.

This is a key point. The standard Bianchi identity for $B$ should have read $dH=0$ but because of the $\al$ corrections, there are both gravitational sources and gauge-field sources of charge:
\be \label{bianchi}
d {H} = { \alpha' \over 4}  \left[\tr (R(\Omega_+) \wedge R(\Omega_+)) - \tr (F \wedge F)  \right].
\ee
It is these sources of NS$5$-brane charge that are concomitant with the the possible violation of the strong energy condition. These sources, together with the $\tr |
R_+|^2$ couplings in the action~\C{hetaction}, play the role of orientifolds in the heterotic string.

There are two natural expansions of the string effective action: the  first is an expansion in space-time derivatives, and the second is an expansion in $\al$. We are retaining all terms up to $O(\al)$ in~\C{hetaction}.
However, with respect to the derivative expansion, we are retaining all $2$ derivative couplings and a special set of the $4$ derivative couplings. Because the heterotic gauge-field kinetic terms are normalized with an $\al$, we are missing couplings like $F^4$ which are $4$ derivatives but order $(\al)^3$. Those terms are certainly interesting for models like gauge-field inflation~\cite{Maleknejad:2011sq}, but we will omit them in this study. Our expansion parameter will be $\al$.

\subsection{Equations of motion}

The supersymmetrization of
the $O(\al)$ interactions, including the $R^2$ and the Lorentz
Chern-Simons couplings, has been worked out with various choices of
fields in~\cite{Chemissany:2007he, Bergshoeff:1989de,
Bergshoeff:1988nn, Metsaev:1987zx}. Ignoring fermions, the equations of motion arising
from this action are
\bea \label{eom}
 R-4 (\nabla \phi)^2+4 \nabla^2 \phi -{1\over 2 } | {
H}|^2-{\alpha'\over 4} \left( \tr | { F} |^2 - \tr |
R_+|^2\right) &=& O(\alpha'^2), \cr
 R_{MN}+2 \nabla_M \nabla_N
\phi -{1\over 4} {H}_{MAB} {{H}_N}^{AB}
-{\alpha'\over 4} \Big[\tr F_{MP}{F_N}^P && \\  -R_{MPAB}(\Omega_+)
R_{N}^{~~PAB}(\Omega_+)\Big] &=& O(\alpha'^2), \cr
 d \left( e^{-2
\phi} \star {H}\right) &=& O(\alpha'^2), \cr
 e^{2\phi} d (e^{-2 \phi} \star {F})+ {A} \wedge \star {F} -
\star {F} \wedge {A} + {F} \wedge \star {
H}& =& O(\alpha'^2). \non
\eea
The dilaton equation of motion has been used to simplify the
Einstein equation appearing above.

To find these equations,
it is easiest to compute the variation of the action with respect to the fields
$\phi$, $g_{MN}$, $B_{MN}$, $A_M$ appearing explicitly, and then
the variation with respect the the connection $\Omega_+$, which
implicitly also depends on these variables. According to a
lemma of~\cite{Bergshoeff:1989de}, the variation of the
$\alpha'$ correction to the action with respect to $\Omega_+$ is
proportional to the leading order equations of motion, and therefore
does not modify the equations of motion to this order.

These results are unique at this order modulo field redefinitions. As long as the action agrees with results from string scattering computations (as checked most recently in~\cite{Chemissany:2007he}), it is determined by supersymmetry. This is known to be true also including terms of $O(\alpha'^2)$. There are no bosonic couplings at order $(\al)^2$ though there are fermionic couplings and possibly modifications to the definition of connections like~\C{conn}.

\subsection{Einstein frame physics}

Define the Einstein frame metric $\tilde{g}_{MN}$ by conformally transforming the string frame metric,
\begin{equation}
  \tilde{g}_{MN}= e^{-\phi/2}g_{MN}.
\end{equation}
The inverse Einstein frame metric is then $\tilde{g}^{MN}=e^{\phi/2}g^{MN}$.  The action becomes
\begin{equation}
  \begin{split}
    S={1\over 2\kappa^2} \int d^{10}x \sqrt{-\tilde{g}} \, & \Big[\tilde{R}
    -\frac{1}{2}\tilde{g}^{MN} \partial_M \phi\partial_N\phi-{1\over 2} e^{-\phi}| {H} |^2 \cr & - {\alpha' \over
      4} e^{-\phi/2}\left( \tr | {F}|^2 -\tr | R_+|^2\right) +O(\alpha'^2) \Big],
  \end{split}
\end{equation}
In expressions involving the Einstein frame metric,
indices will always be raised and lowered with the Einstein frame metric.
In particular, the norms $|\cdot|$ above are taken with respect to the
Einstein frame metric, and this accounts for the additional powers of the
conformal factor.

The Einstein frame equations of motion are
\begin{eqnarray}
\label{dileq}  \tilde{\nabla}^M\tilde{\nabla}_M\phi+\frac{1}{2}e^{-\phi}|H|^2+\frac{\alpha'}{8}e^{-\phi/2}\left(\tr|F|^2-\tr|R_+|^2\right)&=&O(\alpha'^2),\\
\label{einsteq}  \tilde{R}_{MN}-\frac{1}{2}\tilde{g}_{MN}\tilde{R}-\frac{1}{2}\tilde{\nabla}_M\phi\tilde{\nabla}_N\phi+\frac{1}{4}\tilde{g}_{MN}\tilde{\nabla}^P\phi\tilde{\nabla}_P\phi&&\nonumber\\
  -\frac{1}{4}e^{-\phi}H_{MPQ}H_N^{\phantom{M}PQ}+\frac{1}{4}e^{-\phi}\tilde{g}_{MN}|H|^2-\frac{\alpha'}{4}e^{-\phi/2}\Big[\tr F_{MP}F_N^{\phantom{N}P}&&\nonumber\\
  -\frac{1}{2}\tilde{g}_{MN}\tr|F|^2-R_{MPAB}(\Omega_+)R_N^{\phantom{N}PAB}(\Omega_+)+\frac{1}{2}\tilde{g}_{MN}\tr|R_+|^2\Big]&=&O(\alpha'^2),\\
  d\left(e^{-\phi}\tilde{\star}H\right)&=&O(\alpha'^2),\\
  e^{\phi/2} d (e^{-\phi/2} \tilde{\star} {F})+ {A} \wedge \tilde{\star} {F} -
  \tilde{\star} {F} \wedge {A} + {F} \wedge \tilde{\star} {
    H}& =& O(\alpha'^2).
\end{eqnarray}
The Einstein equation can be re-written in Ricci form:
\begin{eqnarray} \label{RicciForm}
  \tilde{R}_{MN}&=&\frac{1}{2}\tilde{\nabla}_M\phi\tilde{\nabla}_N\phi+\frac{1}{4}e^{-\phi}H_{MPQ}H_N^{\phantom{M}PQ}-\frac{1}{8}e^{-\phi}\tilde{g}_{MN}|H|^2+\frac{\alpha'}{4}e^{-\phi/2}\Big[\tr F_{MP}F_N^{\phantom{N}P}\nonumber\\ &&-\frac{1}{8}\tilde{g}_{MN}\tr|F|^2-R_{MPAB}(\Omega_+)R_N^{\phantom{N}PAB}(\Omega_+)+\frac{1}{8}\tilde{g}_{MN}\tr|R_+|^2\Big]+O(\alpha'^2).
\end{eqnarray}

Working in an orthonormal basis, it is easy to see that the
contributions from $\phi$, $H$, and $F$ to $\tilde{R}_{00}$ are
non-negative.  For instance, the time-time contribution of $F$ is
\begin{eqnarray}
  &&\frac{\alpha'}{4}e^{-\phi/2}\tr\left[F_{0I}F_0^{\phantom{0}I}+\frac{1}{16}\left(2F_{0I}F^{0I}+F_{IJ}F^{IJ}\right)\right]\nonumber\\
  &=&\frac{\alpha'}{4}e^{-\phi/2}\tr\left[\frac{7}{8}F_{0I}F_0^{\phantom{0}I}+\frac{1}{16}F_{IJ}F^{IJ}\right]\nonumber\\
  &\ge&0,
\end{eqnarray}
since all terms are positive definite.  Thus, if we were not including
$R_+$ terms in the action, the stress energy tensor would obey the
SEC: 
\be
\tilde{R}_{00} \geq 0. 
\ee 
However, including the $R_+$ terms does not {\it automatically} imply a violation of the SEC. A similar analysis shows that the $R_+$ terms contribute to $R_{00}$ in the following way, 
\bea
&&-{\ap\over4}e^{-\f/2}\left[2R_{+\,0I0J}R_{+\,0}{}^{I0J} + R_{+\,0IJK} R_{+\,0}{}^{IJK} { \over }\right. \non\\
&&+\left.{1\over16}\left(4 R_{+\,0I0J} R_{+}{}^{0I0J} + 2 R_{+\,0IJK} R_{+}{}^{0IJK} + 2 R_{+\,IJ0K} R_{+}{}^{IJ0K} +R_{+\,IJKL}R_{+}{}^{IJKL}\right)\right] \non\\
&=&{\ap\over4}e^{-\f/2}\left[{1\over4}\left(7 R_{+\,0I0J} R_{+\,0}{}^I{}_0{}^J +\hlf R_{+\,IJ0K} R_{+}{}^{IJ}{}_0{}^K\right) \right. \non\\
&&\qquad\quad-\left.{1\over8}\left(7 R_{+\,0IJK} R_{+\,0}{}^{IJK} +\hlf R_{+\,IJKL} R_{+}{}^{IJKL} \right) \right],
\eea
which can be of either sign depending on which terms dominate. To get a feel for when this contribution can become negative, we will need to look more closely at these curvatures. 

\subsection{Evaluating curvatures}

The next step is therefore to evaluate the curvatures, $R_+$, to determine whether acceleration is possible. For useful references, see~\cite{Sen:1986mg, Becker:2009zx}. The usual Christoffel connections are determined by
\be
\G_{IK}^J = {1\over 2} g^{JN} \left( g_{NK, I} + g_{NI, K} - g_{IK,N} \right).
\ee
The inclusion of torsion involves the modification
\be
\G_{IK}^J \, \rightarrow \, \G_{IK}^J - {1\over 2} H_{IK}^{\phantom{IK} J}.
\ee
We need to compute the Riemann tensor with this connection. The background metric is given in~\C{wmetric}. What other background fields, compatible with an FLRW universe, might be excited? In principle, there could  be time-dependence in the dilaton, $\phi$, along with a time-dependent space-time $H$ proportional to the spatial volume form for the FLRW space-time:
\be \label{defh}
H = h(y,t) \e_{ijk} dx^i dx^j dx^k.
\ee
Whether $h$ can depend on $y$ will depend on the form of contributions to the Bianchi identity~\C{bianchi}. These are the universal scalar modes of any heterotic compactification. There can be additional model specific scalar modes that could also be time-dependent.  We will typically simplify equations by setting $h=0$, taking $\f$ to be time-independent, and assuming no additional time-dependent scalars.

Note that the $R_+$ terms appearing on the right hand side of~\C{RicciForm}\ should be computed with the string frame metric, so the metric~\C{wmetric}\ should be multiplied by  $e^{\f/2}$ as in~\C{metricrel}. Under our assumptions about the scalar fields, the non-vanishing components of $R_+$ are just
\bea
R_{+\, \m\n\la}{}^\rho &=& \hat{R}_{\m\n\la}{}^\rho + 2 \dd^\rho_{[\m}\hg_{\n]\la}^{}  |\hn_m\omega|^2, \\
R_{+\, \m m \n}{}^n &=& \hg_{\m\n}\left(X_m{}^n -\hlf e^{-2\omega} H_m{}^{np}\hn_p\omega \right), \\
R_{+\, mnp}{}^q &=& \hat{R}_{mnp}{}^q -2\dd^q_{[m}X_{n]p}^{}-2X^q{}_{[m}\hg_{n]p}-2\dd^q_{[m}\hg_{n]p}^{} |\hn_r\omega|^2 \non\\
&&+e^{-2\omega}\left[\hn_{[m}H_{n]p}{}^q - 2\hn_{[m}\omega H_{n]p}{}^q -H_{mnp}\hn^q\omega + H_{mn}{}^q\hn_p\omega \right] \label{Rinternal}\\
&&+e^{-2\omega}\left(\dd^q_{[m}H_{n]p}^{}{}^r - \hg_{p[m}H_{n]}{}^{qr}\right)\hn_r\omega + \hlf e^{-4\omega} H_{pr[m}H_{n]}{}^{rq} \non,
\eea
where indices are raised and lowered using the product frame metric defined in section~\ref{metricconventions}, and
\be
X_{mn}(\omega) = \hn_m\omega\hn_n\omega - \hn_m\hn_n\omega - \hg_{mn}|\hn_p\omega|^2.
\ee
The expression~\C{Rinternal}\ is rather unwieldy, but fortunately we will never require its explicit form. We have included it here only for completeness.

While these curvature components have been computed from the string frame metric, in~\C{RicciForm}\ they are contracted using the Einstein frame metric. Namely,
\be
\left(R_+^2\right)_{MN} = R_{+\, MPAB}R_{+\,N}{}^{PAB} = W^{-2}R_{+\, MPR}{}^Q R_{+\, NST}{}^U \hg^{PS}\hg^{RT}\hg_{QU}.
\ee
So we find
\bea
\left(R_+^2\right)_{\m\n} 
&=& W^{-2} \left[\hat{R}_{\m\rho\la}{}^\s \hat{R}_{\n}{}^{\rho\la}{}_\s - 4\hat{R}_{\m\n} |\hn_m\omega|^2 \right.\\
&& \left.+ 2\hg_{\m\n}\left( 3\left(|\hn_m\omega|^2\right)^2 + 2|X_{mn}|^2+{1\over2}e^{-4\omega}|H_{mn}{}^p\hn_p\omega|^2 \right)\right], \non\\
\left(R_+^2\right)_{mn} 
&=& W^{-2}\left[R_{+\, m pq}{}^r R_{+\, n}{}^{pq}{}_r {\over }^{} \right.\\
&& + \left. 8\left(X_{mp}X_n{}^p+e^{-2\omega}\hn_p\hn_{(m}\omega H_{n)}{}^{pq}\hn_q\omega +{1\over4} e^{-4\omega}H_{mp}{}^qH_n{}^{pr}\hn_q\omega\hn_r\omega\right)\right], \non
\eea
and the mixed terms $(R_+^2)_{\m m}$ vanish. Note that
\be
|X_{mn}|^2 = {5\over2}\left(|\hn_m\omega|^2\right)^2 + \hn^m\hn_m\omega|\hn_n\omega|^2 +\hlf \hn_m\hn_n\omega\hn^m\hn^n\omega - \hn_m\omega\hn_n\omega\hn^m\hn^n\omega.
\ee
The trace of the curvature squared is given by
\bea
\tr |R_+|^2 &=& \hlf W^{-2}\hg^{MN}(R_+^2)_{MN}  \non\\
&=& W^{-4}\left[ \hlf \hat{R}_{\m\n\rho\s} \hat{R}^{\m\n\rho\s} - 2 \hg^{\m\n}\hat{R}_{\m\n}|\hn_m\omega|^2 + \hlf R_{+\, mnpq} R_{+\,}{}^{mnpq} \right.\\
&& \left. +4\left(3\left(|\hn_m\omega|^2\right)^2 + 4|X_{mn}|^2 + e^{-4\omega}|H_{mn}{}^p\hn_p\omega|^2\right) \right] \non.
\eea

Using the curvature components of a spatially flat FLRW metric we can now compute the total contribution of $R_+$ to the right hand side of the time-time component of the Einstein equation~\C{RicciForm}:
\bea
&&-{\ap\over4}e^{-\f/2}\left[(R_+^2)_{00} -{1\over8}\tg_{00} \tr|R_+|^2\right] \non\\
&=& {\ap\over4}e^{-2\omega}\left[{3\over4}\left(7\left({\ddot{a}\over a}\right)^2-\left({\dot{a}\over a}\right)^4 \right) -{3\over2}\left(7\left({\ddot{a}\over a}\right)- \left({\dot{a}\over a}\right)^2\right)\left|\hn_m\omega\right|^2  \right. \\
&&\left. +{9\over2}\left(|\hn_m\omega|^2\right)^2 +  2 |X_{mn}|^2 +\hlf e^{-4\omega}|H_{mn}{}^p\hn_p\omega|^2 -{1\over16} R_{+\, mnp}{}^q R_{+\,}{}^{mnp}{}_q \right] \non.
\eea
Extending this result to non-spatially flat FLRW metrics is straightforward, simply replace $\dot{a}^2$ everywhere with $\dot{a}^2 + k$.

\section{Studying the Stringy Equations of Motion}

As noted earlier, with the addition of $O(\alpha')$ corrections, the
time-time component of the Einstein equation in Ricci form~\eqref{RicciForm}\ is not positive definite. Therefore the SEC can be violated, and we can potentially avoid the supergravity no-go theorem.  On the other
hand, we must satisfy the dilaton equation \eqref{dileq}.  This
effectively puts a bound on the amount of SEC violation that can
occur, and we can make use of this bound by substituting \eqref{dileq}
into \eqref{RicciForm},
\bea
\tilde{R}_{MN} &=& {1\over4}\tilde{g}_{MN}\tilde{\nabla}^P\tilde{\nabla}_P\f + \hlf \tilde{\nabla}_M\f\tilde{\nabla}_N\f + {1\over4}e^{-\f} H_{MPQ}H_N{}^{PQ} \non\\
&&+{\ap\over4}e^{-\f/2}\left[\tr F_{MP}F_N{}^P -R_{+\, MPAB}R_{+\,N}{}^{PAB}\right] .
\eea
This can be conveniently written using the product frame metric~\C{metricrel}\ (with no assumptions on the dilaton or $h$) as,
\bea
\hat{R}_{MN} -\hg_{MN}\left(\hn^\m\hn_\m\omega + W^{-8}\hn^m(W^8\hn_m\omega)\right) + 8W\hn_M\hn_N W^{-1} \non\\
= \hlf\hn_M\f\hn_N\f + {1\over4}e^{-2\omega}H_{MPQ}H_N{}^{PQ} +{\ap\over4}e^{-2\omega}\left[\tr F_{MP}F_N{}^P - \left(R_+^2\right)_{MN}\right],
\eea
where we have moved the $\tn^2\f$ to the left hand side and combined it with the warp factor. In the expression above, and those that follow, indices are
raised and lowered with $\hat{g}_{MN}$.  
When both $M,N$ lie in space-time we find:
\bea\label{Rmunu}
\hat{R}_{\m\n}-\hg_{\m\n}\left(\hn^\la\hn_\la\omega + W^{-8}\hn^m(W^8\hn_m\omega)\right)  &=& \hlf\hn_\m\f\hn_\n\f + {1\over2}e^{-4\omega}h^2(\hg_{\m\n}+n_\m n_\n ) \nonumber \\* && -{\ap\over4}e^{-2\omega}\left(R_+^2\right)_{\m\n}.
\eea 
The vector $n_\mu$ is the unit normal to the homogeneous space-like hypersurface, defined in~\C{definen}. The case with one index internal and one in space-time is rather interesting:
\be
0=\hn_\m\f \hn_m\f - {\ap\over2}e^{-2\omega}\left(R_+^2\right)_{\m m}.
\ee
Although this is trivially satisfied when $\f$ is static, it should impose interesting constraints on the allowed dilaton time dependence. Finally, the case with both indices internal is not terribly interesting since no terms drop out. It provides a constraint on the the compactification manifold and internal fluxes. For completeness, we write it anyway:
\bea
\hat{R}_{mn} -\hg_{mn}\left(\hn^\m\hn_\m\omega + W^{-8}\hn^p(W^8\hn_p\omega)\right) + 8W\hn_m\hn_n W^{-1}
= \hlf\hn_m\f\hn_n\f  \non\\ + {1\over4}e^{-2\omega}H_{mpq}H_n{}^{pq} +{\ap\over4}e^{-2\omega}\left[\tr F_{mp}F_n{}^p - \left(R_+^2\right)_{mn}\right].  \label{Einsteqint}
\eea
Under the assumption of static $\f$ and $h=0$, we take $\omega=\omega(y)$. Equation~\C{Rmunu}\ then reduces to
\bea\label{ddomega}
\hat{g}_{\mu\nu}W^{-8}\hat{g}^{mn}\hat{\nabla}_m\left(W^8\hat{\nabla}_n\omega\right)  =\hat{R}_{\m\n} +{\ap\over4}e^{-2\omega}\left[\hat{R}_{\m\rho\la}{}^\s \hat{R}_{\n}{}^{\rho\la}{}_\s - 4\hat{R}_{\m\n} |\hn_m\omega|^2 \right.\\
\left.+ 2\hg_{\m\n}\left( 3\left(|\hn_m\omega|^2\right)^2 +
    2|X_{mn}|^2+{1\over2}e^{-4\omega}|H_{mn}{}^p\hn_p\omega|^2 \right)\right].
\non \eea

\subsection{Minkowski Space-time}
For Minkowski space-time,~\eqref{ddomega}\ reduces to the simple constraint
\bea
\hn^m\left(W^8\hn_m\omega\right)  ={\ap\over2}e^{-2\omega}W^{8}\left[3\left(|\hn_m\omega|^2\right)^2 + 2|X_{mn}|^2+{1\over2}e^{-4\omega}|H_{mn}{}^p\hn_p\omega|^2 \right]
\eea
Integrating this over the internal space, we see that the left hand side vanishes. Composed of a sum of squares, the right hand side must then vanish identically. This implies that $\omega$ is constant. We conclude that for any background with a Minkowski space-time the string frame metric must be unwarped. Any warping that appears in Einstein frame can be identified with the dilaton. Supersymmetric torsional compactifications are special solutions of this type.

\subsection{(A)dS Space-time}

One can extend this analysis to maximally symmetric space-times with cosmological constant  to find that $\Lambda>0$ is not allowed,
but $\La<0$ is possible.  To see this, substitute
\be
\hat{R}_{\m\n\la\rho} = {2\La\over3}\, \hg_{\la[\m}\hg_{\n]\rho}
\ee
into equation \eqref{ddomega}\ yielding
\bea
 W^{-8} \hn^m\left(W^8\hn_m\omega\right) &=& \Lambda + 
{\ap\over2}e^{-2\omega}\Big[\frac{1}{3}\left(\Lambda-3|\hn_m\omega|^2\right)^2  +
  2|X_{mn}|^2 \cr && +{1\over2}e^{-4\omega}|H_{mn}{}^p\hn_p\omega|^2 \Big]. 
\eea
Note that the terms involving the space-time Riemann tensor and
$|\hat{\nabla}_m\omega|^2$ combined into a perfect square!

As with the previously considered Minkowski case, we can integrate
over the internal space, which causes the left hand side to vanish.
We are left with 
\be \La = -{\ap\over2V'}\int_\cM d^6y
\sqrt{\hg_6} W^6 e^{-\f/2}\bigg[3|\hn_m\omega|^2 + 2|X_{mn}|^2
  +{1\over2}e^{-4\omega}|H_{mn}{}^p\hn_p\omega|^2 \bigg]  + O(\ap^2),  \ee
  where $V'=\int d^6 y\sqrt{\hg_6} W^8$.
  The positive-definiteness of the integrand tells us that $\Lambda\le0$, so
that a dS space-time is ruled out.

On the other hand, a small negative cosmological constant (AdS) is
allowed, as an $O(\alpha')$ effect.  Here, $\La<0$ acts as a sink for
$\f$.  These weakly curved heterotic AdS solutions are classically
forbidden and only arise as a result of the $\tr|R_+|^2$ correction.
One should note, however, that even though these no-go arguments fail
for AdS, it does not mean that such a solution necessarily exists since one must
solve the complete set of equations of motion. This, however, should be possible to do starting with a non-supersymmetric type IIB flux compactification on $K3\times T^2$ 
and dualizing to heterotic following~\cite{Dasgupta:1999ss}.  Indeed this dualization has been 
performed in~\cite{Becker:2009zx}\ at the level of supergravity; it would be very interesting to see if the resulting solution (including $\ap$ corrections) is actually AdS$_4$. 

\subsection{General FLRW Space-time}

It is also of interest to investigate more general FLRW space-times, with potential applications to inflation.  A complete investigation of this case requires dynamical scalars resulting in a considerably more complicated (and model-dependent) analysis. That case will be explored elsewhere.    


Since our space-time is no longer maximally symmetric, but has FLRW symmetry, we must consider separately the time-time and space-space components of~\C{ddomega}. The time-time component is
\bea\label{milne}
W^{-8} \hn^m\left(W^8\hn_m\omega\right) &=& 3\left({\ddot{a}\over a}\right) + {\ap\over2}e^{-2\omega}\bigg[3\left({\ddot{a}\over a} - |\hn_m\omega|^2\right)^2 + 2|X_{mn}|^2 \nonumber \\* && + \hlf e^{-4\omega}|H_{mn}{}^p\hn_p\omega|^2\bigg],
\eea
while the space-space components give
\bea
W^{-8} \hn^m\left(W^8\hn_m\omega\right) &=& \left({\ddot{a}\over a}\right) + 2\left({\dot{a}^2+k\over a^2}\right) + {\ap\over2}e^{-2\omega}\left[\left({\ddot{a}\over a}-|\hn^m\omega|^2\right)^2 \right. \\ &&
\left. + 2\left({\dot{a}^2+k\over a^2} - |\hn_m\omega|^2\right)^2 + 2|X_{mn}|^2 + \hlf e^{-4\omega}|H_{mn}{}^p\hn_p\omega|^2  \right]\non.
\eea
Let us  integrate both of these equations over the internal space-time.  As before, the left hand sides both vanish, and we are left with the equivalent of the Friedmann equations within this setup.

To $O(\alpha')$, the source for these Friedmann equations is simply an
effective cosmological constant, 
\begin{eqnarray} \La_{\text{eff}}&\equiv&
-{\ap\over 2V'}\int_\cM d^6y \sqrt{\hg_{6}} W^6
e^{-\f/2}\left[3|\hn_m\omega|^2 + 2|X_{mn}|^2
  +{1\over2}e^{-4\omega}|H_{mn}{}^p\hn_p\omega|^2 \right] + O(\ap^2) \nonumber\\
&\le&0.
\end{eqnarray} 
This gives nothing new beyond what we have seen before.  Indeed, in the case
where $\omega=0$, and hence $\Lambda_{\text{eff}}=0$, we can solve to
obtain $\dot{a}=-k=1$. This is the Milne universe, which is (part of)
Minkowski space-time, which we already found.

In the case where $\Lambda_{\text{eff}}<0$, we can solve the effective
Friedmann equations to obtain
\begin{eqnarray}
  a(t)&=&\sin\left(\sqrt{-\frac{\Lambda_{\text{eff}}}{3}}t\right)\,,\\
  k&=&\frac{\Lambda_{\text{eff}}}{3}\,.
\end{eqnarray}
While this looks slightly more non-trivial, this solution actually
represents a patch of AdS, which, again, was examined in the previous
section.  To see this, we note that for FLRW metrics, the Weyl tensor
vanishes, so the Riemann tensor can be written entirely in terms of
the Ricci tensor.  This, in turn, can be written in terms of the Ricci
scalar.  Putting these observations together,
\begin{eqnarray}
  \hat{R}_{\mu\nu\lambda\rho}&=&\left(\hat{g}_{\mu[\lambda}\hat{R}_{\rho]\nu}-\hat{g}_{\nu[\lambda}\hat{R}_{\rho]\mu}\right)-\frac{1}{3}\hat{R}\hat{g}_{mu[\lambda}\hat{g}_{\rho]\nu}, \nonumber\\
  &=&\frac{2\Lambda_{\text{eff}}}{3}\hat{g}_{mu[\lambda}\hat{g}_{\rho]\nu}\,,
\end{eqnarray}
so the space-time is maximally symmetric, and hence (part of) AdS. In summary, by considering more general FLRW space-times, we do not find any additional solutions.

\subsection*{Acknowledgements}

It is our pleasure to thank Andy Albrecht, Andrew Ulvestad and Bob Wald for discussions. S.~R.~G. is supported in part by NSF Grant No. PHY08-54807. E.~M. is supported in part by DOE grant DE-FG02-90ER-40560. C.~Q. and S.~S. are supported in part by
NSF Grant No.~PHY-0758029 and NSF Grant No.~0529954.

\appendix
\section{Conventions}
\label{conventions}
Here we summarize our conventions, and list several useful formulae. We will provide more general formulae than are needed in this analysis by allowing time-dependence for the dilaton and for $h$ defined in~\C{defh}. The following table summarizes the different indices we use throughout the paper.
\begin{center}
\begin{tabular}{ll}
\hline
Letters     & Use for: \\
\hline
$A,B,\ldots$ & 10-dimensional local Lorentz \\
$M,N,\ldots$ & 10-dimensional space-time \\
$I,J,\ldots$ & ~9-dimensional space \\
$\m,\n,\ldots$ & ~4-dimensional space-time \\
$i,j,k,\ell$ & ~3-dimensional space \\
$m,n,\ldots$ & ~6-dimensional internal space \\
\hline
\\
\end{tabular}
\end{center}
The norm of a rank $p$ tensor is defined by
\be
|T_{(p)}|^2 = {1\over p!} g^{M_1 N_1}\ldots g^{M_pN_p} T_{M_1\ldots M_p}T_{N_1\ldots N_p}.
\ee
When (anti-)symmetrizing indices on tensors, we normalize them with an overall $1/(p!)$, so for example
\be
T_{[M_1\ldots M_p]} ={1\over p!}\sum_\s (-1)^{|\s|} T_{M_{\s(1)}\ldots M_{\s(p)}}
\ee
where $|\s|$ denotes the order of the permutation $\{1,2,\dots, p\}\rightarrow\{\s(1),\s(2),\ldots,\s(p)\}$. In particular, $T_{[MN]} = \hlf(T_{MN} - T_{NM})$.

\subsection{Computing Curvatures}
The Riemann tensor is defined by
\be \label{Riemann}
R_{MNP}{}^Q = 2\del_{[N}^{}\G^Q_{M]P} + 2 \G^Q_{R[N}\G^R_{M]P}.
\ee
Under a conformal transformation of the metric,
\be
g_{MN} = e^{2\omega}\hg_{MN},
\ee
the Christoffel connection shifts to
\be
\G^P_{MN} = \hat{\G}^P_{MN} + C^P_{MN}
\ee
where
\be
C^P_{MN} = 2\dd^P_{(M}\hn_{N)}^{}\omega - \hg_{MN}\hn^P\omega,
\ee
and hatted quantities are built using the metric $\hg$, which is also used to raise and lower indices. In terms of the $\hg$, the curvature of the conformally transformed metric is
\bea
R_{MNP}{}^Q &=& \hat{R}_{MNP}{}^Q - 2\hn_{[M}^{}C_{N]P}^Q + 2C^R_{P[M}C^Q_{N]R}, \\
&=& \hat{R}_{MNP}{}^Q + 2\dd^Q_{[M}\hn_{N]}^{}\hn_P^{}\omega - 2\hg_{P[M}\hn_{N]}\hn^Q\omega \non\\
&& + 2\hn_{[M}^{}\omega\dd_{N]}^Q \hn_P\omega - 2\hn_{[M}^{}\omega\hg_{N]P}^{}\hn^Q\omega + 2\dd^Q_{[M}\hg_{N]P}^{} \left|\hn_R\omega\right|^2 \non.
\eea
Then by contraction we get the conformally transformed Ricci tensor and scalar:
\bea
R_{MN} &=& \hat{R}_{MN} - \hg_{MN} \hn^2\omega + (D-2) \bigg(\hn_M\omega\hn_N\omega - \hn_M\hn_N\omega -\hg_{MN}\left|\hn_P\omega\right|^2\bigg), \\
R &=& e^{-2\omega}\left(\hat{R} - 2(D-1)\hn^2\omega - (D-2)(D-1)\left|\hn_M\omega\right|^2\right).
\eea
Another useful relation is the behaviour of the scalar wave operator under a conformal transformation:
\be
\nabla^2 = e^{-2\omega}\left(\hn^2 + (D-2)\hn^M\omega\hn_M \right).
\ee

\subsection{Curvature with Torsion}
Torsion shows up in our discussion via the modified spin connection,
\be
\omega_\pm^A{}_{BM} = \omega^A{}_{BM} \pm \hlf H^A{}_{BM}.
\ee
This leads to the torsionful affine connection:
\bea
\G^P_{\pm\, MN} = e^P_A\left(\del_M e^A_N + e^B_N\omega_\pm^A{}_{BM}\right) = \G^P_{MN} \mp\hlf H^P{}_{MN}.
\eea
The fact that $H$ is totally anti-symmetric implies that these connections are still compatible with the metric. The curvatures, $R_\pm$, of these connections are given by the same expression~\C{Riemann}\ as before, except now we replace the standard Christoffel connection with the torsionful one from above. In particular, this leads to
\bea
R_{+\, MNP}{}^Q = R_{MNP}{}^Q + \nabla_{[M}^{}H_{N]P}{}^Q + \hlf H_{PR[M}^{}H_{N]}^{\phantom{N}RQ}.
\eea
It is useful to note that under a conformal transformation, the second term in the line above gets modified as follows
\bea
\nabla_{[M}H_{N]P}{}^Q &=& \hn_{[M}\left(g^{QR}H_{N]PR}\right) + C^Q_{R[M}H_{N]P}^{}{}^R - C^R_{P[M}H_{N]R}^{}{}^Q, \\
&=& e^{-2\omega}\left[\hn_{[M}H_{N]P}{}^Q - 2\hn_{[M}^{}\omega H_{N]P}^{\phantom{N]P}Q} - H_{MNP}\hn^Q\omega + H_{MN}{}^Q\hn_P\omega \right. \non\\
&&\qquad +\left.\left(\dd^Q_{[M}H_{N]P}^{\phantom{N]P}R} - \hg_{P[M}H_{N]}{}^{QR}\right)\hn_R\omega\right] \non,
\eea
while the $H^2$ term picks up a factor of $e^{-2\omega}$.

Many useful facts and formulae about torsionful connections have been worked out by S. Jensen, and can be found in~\cite{Jensennotes}. We will summarize those that are pertinent to our work.
The symmetry properties of $R_+$ are slightly different from the standard case. With all indices lowered, we still have anti-symmetry in the first and last pairs:
\be
R_{+\, MNPQ} = R_{+\, [MN][PQ]}.
\ee
However, under interchange of the first pair with the last pair, we now have
\be
R_{+\, MNPQ} = R_{-\, PQMN} - 2\del_{[M}H_{NPQ]}, \label{R+symm}
\ee
where $R_-$ is the curvature of $\G_-$. In a heterotic string background, $dH\sim O(\ap)$ and so we can set it to zero in computing $R_+^2$, since those curvature terms appear in the action already at $O(\ap)$.

\subsection{FLRW Space-times}
The metric of an FLRW space-time takes the form,
\be
ds^2 = -dt^2 + a^2(t) h_{ij}(x)dx^i dx^j, \label{FLRW}
\ee
where the spatial part of the metric can be written
\be
h_{ij} = \dd_{ij} + k {x_i x_j \over 1-k x^2}.
\ee
The non-vanishing components of the (torsion-free) connection are
\bea
\G_{0j}^i = \left({\dot{a}\over a}\right) \dd^i_j, \qquad \G^0_{ij} = a\dot{a} h_{ij},\qquad \G^k_{ij} = \hlf h^{k\ell}\left(2\del_{(i} h_{j)\ell} - \del_\ell h_{ij}\right),
\eea
and the only non-trivial components of the Riemann curvature are
\bea
R_{i0j}{}^0 = a\ddot{a}\, h_{ij}, \qquad R_{ijk}{}^\ell = 2\left(k +(\dot{a})^2\right)\dd^\ell_{[j}h_{i]k}^{}.
\eea

The only way to turn on torsion in such a space-time, while preserving homogeneity and isotropy, is if $H$ takes the form
\be \label{definen}
H = {1\over 3!}h(t)n^\s \e_{\s\m\n\la} dx^\m dx^\n dx^\la
\ee
where $n^\m = \dd^\m_0$ is the unit vector normal to the homogeneous space-like hypersurface. $h$ cannot be an arbitrary function of $t$, since it must satisfy the Bianchi identity:
\be
\dot{h} + 3 \left({\dot{a} \over a} \right) h = {\ap\over4}\left(\ldots\right) .
\ee
It will be useful later to note that,
\be
\nabla_\m n^\n = \G^\n_{\m 0} = \left({\dot{a}\over a}\right)\left(\dd^\n_\m + n^\n n_\m\right),
\ee
and the $3$-volume form $\e_{ijk}$ is the one associated with $g_{ij}$, not $h_{ij}$, so
\be
\del_t \e_{ijk} = 3\left({\dot{a}\over a}\right) \e_{ijk}.
\ee

\subsection{FLRW$\times_W \, \cM$}
When we consider a warped product space-time of the form FLRW times some compact internal manifold $\cM$, then the most general ansatz we can allow in string frame is
\bea
ds^2 &=& e^{2\omega(t,y)}\left(\hg_{\m\n}(x)dx^\m dx^\n + \hg_{mn}(y)dy^m dy^n\right), \\
H &=& {1\over 3!}\left( h(t,y) \e_{\s\m\n\la}n^\s dx^\m dx^\n dx^\la + H_{mnp}(y) dy^m dy^n dy^p \right).
\eea
The conformal factor $\omega$ is a combination of a warp factor $W$ and the dilaton $\f$
\be
\omega(t,y) = \log W(y) + {1\over4}\f(t,y),
\ee
and the metric $\hg_{\m\n}$ is the usual FLRW metric~\C{FLRW}. The components of curvature associated to the torsionful connection are then
\bea
R_{+\,\m\n\la}{}^\rho &=& \hat{R}_{\m\n\la}{}^\rho + 2\dd^\rho_{[\m}\hn_{\n]}^{}\hn_\la^{}\omega - 2\hg_{\la[\m}\hn_{\n]}\hn^\rho\omega \\
&& + 2\hn_{[\m}^{}\omega\dd_{\n]}^\rho \hn_\la\omega - 2\hn_{[\m}^{}\omega\hg_{\n]\la}^{}\hn^\rho\omega + 2\dd^\rho_{[\m}\hg_{\n]\la}^{} \left(|\hn_\m\omega|^2+|\hn_m\omega|^2\right) \non\\
&& + e^{-2\omega}\left[-\hn_{[\m}\left(h n^\s\right)\e_{\n]\la}{}^\rho{}_{\s} + h n^\s\left(2 \hn_{[\m}\omega\e_{\n]\la}{}^\rho{}_{\s} + \e_{\m\n\la\s}\hn^\rho\omega - \e_{\m\n}{}^{\rho}{}_\s \hn_\la\omega \right)\right] \non\\
&& + \hlf e^{-4\omega} h^2\left[\dd^\rho_{[\m}\hg^{}_{\n]\la} + \dd^\rho_{[\m} n_{\n]}^{} n_\la^{} + n_{[\m}\hg_{\n]\la}n^\rho \right], \non \\
R_{+\, \m\n\la}{}^m &=& -2\hg_{\la[\m}\hn_{\n]}\hn^m\omega + 2\hg_{\la[\m}\hn_{\n]}\omega\hn^m\omega + e^{-2\omega}h\e_{\m\n\la\s}n^\s\hn^m\omega, \\
R_{+\,\m\n m}{}^n &=& 0, \\
R_{+\, \m m\n}{}^n &=& \dd^n_m\left(\hn_\m\omega\hn_\n\omega -\hn_\m\hn_\n\omega\right) + \hg_{\m\n}\left(\hn_m\omega\hn^n\omega -\hn_m\hn^n\omega\right) \\
&&-\dd^n_m\hg_{\m\n} \left(|\hn_\m\omega|^2+|\hn_m\omega|^2\right) -\hlf e^{-2\omega}\hg_{\m\n} H_m{}^{np}\hn_p\omega, \non\\
R_{+\,mnp}{}^\m &=& -2\hg_{p[m}\hn_{n]}\hn^\m\omega + 2\hg_{p[m}\hn_{n]}\hn^\mu\omega - e^{-2\omega} H_{mnp}\hn^\m\omega, \\
R_{+\, mnp}{}^q &=& \hat{R}_{mnp}{}^q +2\dd^q_{[m}\hn_{n]}^{}\hn_p\omega - 2\hg_{p[m}\hn_{n]}\hn^q\omega \non\\
&&-2\dd^q_{[m}\hn_{n]}^{}\omega\hn_p\omega - 2\hn_{[m}^{}\omega\hg_{n]p}^{}\hn^q\omega +2\dd^q_{[m}\hg_{n]p}^{} \left(|\hn_\m\omega|^2+|\hn_m\omega|^2\right) \\
&&+e^{-2\omega}\left[\hn_{[m}H_{n]p}{}^q - 2\hn_{[m}\omega H_{n]p}{}^q -H_{mnp}\hn^q\omega + H_{mn}{}^q\hn_p\omega \right] \non\\
&&+e^{-2\omega}\left(\dd^q_{[m}H_{n]p}^{}{}^r + \hg_{p[m}H_{n]}{}^{qr}\right)\hn_r\omega + \hlf e^{-4\omega} H_{pr[m}H_{n]}{}^{rq} \non.
\eea
After separating out the time and space indices, we can be a little more explicit. Up to symmetries, including~\C{R+symm}, the non-trivial components are
\bea
R_{+\,0i0}{}^j &=& -\dd^j_i\left[\left({\ddot{a}\over a}\right) +  \ddot{\omega} + \left({\dot{a}\over a}\right)\dot{\omega} - |\hn_m\omega|^2  \right], \\
R_{+\,ijk}{}^0 &=&  e^{-2\omega}h\e_{ijk} \left[\dot{\omega} +\left({\dot{a}\over a}\right) \right], \\
R_{+\,ijk}{}^\ell &=& 2a^2\dd^\ell_{[i}h_{j]k} \left[{k\over a^2}+\left({\dot{a}\over a}\right)^2 -2\left({\dot{a}\over a}\right)\dot{\omega} - (\dot{\omega})^2 + |\hn_m\omega|^2 + {1\over4}e^{-4\omega}h^2   \right],\\
\non \\
R_{+\,0ij}{}^m &=& a^2 h_{ij}\left[\hn^m\dot{\omega} - \dot{\omega}\hn^m\omega  \right],\\
R_{+\,ijk}{}^m &=& -e^{-2\omega}h\e_{ijk}\hn^m\omega, \\
\non \\
R_{+\,0m0}{}^n &=& -\dd^n_m\ddot{\omega} -X_m{}^n(\omega) +\hlf e^{-2\omega} H_m{}^{np}\hn_p\omega, \non\\ \\
R_{+\,imj}{}^n &=& a^2h_{ij}\left(\dd_m^n\left({\dot{a}\over a} + \dot{\omega}\right)\dot{\omega} + X_m{}^n(\omega) -\hlf e^{-2\omega}H_m{}^{np}\hn_p\omega \right), \\
\non \\
R_{+\,mnp}{}^0 &=& 2\hg_{p[m}\hn_{n]}\dot{\omega} - \left(2\hg_{p[m}\hn_{n]}\omega -e^{-2\omega}H_{mnp}\right)\dot{\omega},\\
\non\\
R_{+\,mnp}{}^q &=& \hat{R}_{mnp}{}^q +2\dd^q_{[m}\hn_{n]}^{}\hn_p\omega - 2\hg_{p[m}\hn_{n]}\hn^q\omega \non\\
&&-2\dd^q_{[m}\hn_{n]}^{}\omega\hn_p\omega - 2\hn_{[m}^{}\omega\hg_{n]p}^{}\hn^q\omega +2\dd^q_{[m}\hg_{n]p}^{} \left((\dot{\omega})^2+|\hn_r\omega|^2\right) \\
&&+e^{-2\omega}\left[\hn_{[m}H_{n]p}{}^q - 2\hn_{[m}\omega H_{n]p}{}^q -H_{mnp}\hn^q\omega + H_{mn}{}^q\hn_p\omega \right] \non\\
&&+e^{-2\omega}\left(\dd^q_{[m}H_{n]p}^{}{}^r + \hg_{p[m}H_{n]}{}^{qr}\right)\hn_r\omega + \hlf e^{-4\omega} H_{pr[m}H_{n]}{}^{rq} \non.
\eea
In the Einstein equations, the combination $\left(R_+^2\right)_{MN} = W^{-2}R_{+\, MPQ}{}^R R_{+\,N ST}{}^U\hg^{PS}\hg^{QT}\hg_{RU}$ appears. The non-vanishing components are
\bea
\left(R_+^2\right)_{00} &=& W^{-2}\left\{16\left|\hn_m\dom - \dom\hn_m\omega\right|^2 + 6 e^{-4\omega}h^2\left(\dom + {\da\over a}\right)^2 + 6\dom^2 e^{-4\omega}\left|H_{mnp}\right|^2 \right. \\
&&\quad \left.-6\left({\dda\over a}+\ddo + \dom \left({\da\over a}\right) -|\hn_m\omega|^2\right)^2 -4\left|\hg_{mn}\ddo +X_{mn}\right|^2 - e^{-4\omega}\left|H_{mn}{}^p\hn_p\omega\right|^2 \right\}, \non \\
\left(R_+^2\right)_{ij} &=& 2W^{-2}\hg_{ij}\left\{\left({\dda\over a}+\ddo + \dom\left({\da\over a}\right) -|\hn_m\omega|^2\right)^2 +2\left|\hg_{mn}\dom\left(\dom + {\da\over a}\right) + X_{mn}\right|^2 \right.\non\\
&&\quad+ 2\left({k+\da^2\over a^2} + 2\dom\left({\da\over a}\right) + \dom^2 - |\hn_m\omega|^2 - {1\over4}e^{-4\omega}h^2\right)^2  + e^{-4\omega}h^2|\hn_m\omega|^2 \\
&&\quad\left. +\hlf e^{-4\omega}\left|H_{mn}{}^p\hn_p\omega\right|^2 - 3e^{-4\omega}h^2\left(\dom+{\da\over a}\right)^2 - \left|\hn_m\dom -\dom\hn_m\omega\right|^2\right\}, \non \\
\left(R_+^2\right)_{0m}&=& W^{-2}\left\{6e^{-4\omega}h^2\left(\dom + {\da\over a} \right)\hn_m\omega + \dom e^{-4\omega}\left(\hlf H_{mnp}\hn_q\omega - e^{2\omega}R_{+\,mnpq}\right)H^{npq} \right. \non\\
&&\quad -(\hn_n\dom -\dom\hn_n\omega)\left[\dd^n_m\left(6{\dda\over a} + 11\ddo +12\dom\left({\da\over a}\right) + 6\dom^2 - 11|\hn_p\omega|^2 - \hn^p\hn_p\omega \right) \right. \non\\
&&\qquad\qquad\qquad\qquad\quad \left.\left. + 5 X_m{}^n - {7\over2}e^{-2\omega}H_m{}^{np}\hn_p\omega + 2 R_{+\,mp}{}^{np} \right]\right\}, \label{R^2int}\\
\left(R_+^2\right)_{mn}&=& W^{-2}\bigg\{R_{+\,mpq}{}^r R_{+\,n}{}^{pq}{}_r {\over} + 2\left(\hg_{mn}\ddo^2 + 2\ddo X_{mn} + X_m{}^pX_{pn}\right)  + 6e^{-4\omega}h^2\hn_m\omega\hn_n\omega \non\\
& + & \hlf e^{-4\omega} H_{mr}{}^p H_n{}^{rq}\hn_p\omega\hn_q\omega - 6\hg_{mn}\left|\hn_p\dom -\dom\hn_p\omega\right|^2 - 3\dom^2e^{-4\omega} H_{mpq}H_n{}^{pq} \bigg\}. 
\eea
In addition, we will require the fully traced curvature-squared:
\bea
\tr|R_+|^2 &=& \hlf R_{+\,mnpq}R_+{}^{mnpq} + 6\left({\dda\over a}+\ddo + \dom\left({\da\over a}\right) -|\hn_m\omega|^2\right)^2 \non \\
&& + 6 \left({k+\da^2\over a^2}- 2\dom\left({\da\over a}\right) -\dom^2 + |\hn_m\omega|^2 + {1\over4}e^{-4\omega}h^2\right)^2 \non\\
&& + 4 \left|{\over}\hg_{mn}\ddo + X_{mn}\right|^2  + 6  \left|\hg_{mn}\dom\left(\dom + {\da\over a}\right) + X_{mn}\right|^2 \\
&& + 6e^{-4\omega} h^2|\hn_m\omega|^2 +  {5\over2}e^{-4\omega}\left|H_{mn}{}^p\hn_p\omega\right|^2  \non\\
&& - 14 \left|\hn_m\dom -\dom\hn_m\omega\right|^2-12 \dom^2 e^{-4\omega}\left|H_{mnp}\right|^2  - 12 e^{-4\omega}h^2\left(\dom + {\da\over a}\right)^2 \non.
\eea

\newpage

\ifx\undefined\bysame
\newcommand{\bysame}{\leavevmode\hbox to3em{\hrulefill}\,}
\fi

\end{document}